# Increment entropy as a measure of complexity for time series

Xiaofeng Liu[1,2,*], Aimin Jiang[1,2], Ning Xu[1,2], Jianru Xue[3],

1 College of IoT Engineering, Hohai University, Changzhou, 213022, China.
2 Changzhou key laboratory of robotics and intelligent technology, Changzhou, 213022, China
3 Institute of Artificial Intelligence and Robotics, Xi'an Jiaotong University, Xi'an,710049, China
[*] E-mail : Corresponding xfliubme@gmail.com

## Abstract

Entropy has been a common index to quantify the complexity of time series in a variety of fields. Here, we introduce increment entropy to measure the complexity of time series in which each increment is mapped into a word of two letters, one letter corresponding to direction and the other corresponding to magnitude. The Shannon entropy of the words is termed as increment entropy (IncrEn). Simulations on synthetic data and tests on epileptic EEG signals have demonstrated its ability of detecting the abrupt change, regardless of energetic (e.g. spikes or bursts) or structural changes. The computation of IncrEn does not make any assumption on time series and it can be applicable to arbitrary real-world data.

## Author summary

## Introduction

Nowadays, the notion of complexity has been ubiquitously used to examine a variety of time series, ranging from diverse physiological signals [1-9] to financial time series [10, 11] and ecological time series [12]. There there is not an established universal definition of complexity to date [13]. In practice, various measures of complexity have been developed to characterize the behaviors of time series, e.g., regular, chaotic, and stochastic behaviors [8, 14-18]. Among these studies, approximate entropy (ApEn) proposed by Pincus [16] is one of the most commonly used methods. It is a measure of regularity of chaotic and non-stationary time series [19-23]. It has been demonstrated that the more frequent similar epochs occur in time series, the lower the resulting ApEn is. As pointed out by Pincus [16], AnEn indicates the generation rate of new information. Richman and Moorman later proposed the notion of sample entropy (SampEn) [8] that is conceptually inherited from ApEn but computationally counteract the shortcomings of ApEn. SampEn excludes self-matches of vectors and demonstrates relatively consistency [8, 24].

In recent years, Permutation entropy (PE) has been widely applied to quantify the complexity of a variety of experimental time series [25,26], in particular physiological signal [27-32]. PE is robust to observation or dynamical noise, conceptually simple, and computationally extremely fast. These advantages make it suitable for analysing data sets of huge size without any preprocessing and fine-tuning of parameters. An intriguing feature of PE is that a time series is naturally mapped into a symbol sequence by comparing neighboring values. However, this ordinal pattern ignores the relative sizes of neighboring values. Recently Liu and Wang added a factor related to magnitude of consecutive values of time series into an mapping pattern vector, so that the epochs with an identical ordinal pattern can be discriminated [33]. This fine-graining partition makes the modified version of PE more sensitive to abrupt changes in amplitude, such as bursts and spikes [34, 35]. Another modified version of PE, is weighted-permutation entropy, introduced by Fadlallah and colleagues [36]. Similarly, they incorporated amplitude information into the pattern extracted from a given time series. As a result, their new scheme performs better when tracking the abrupt changes in signal. In addition, in PE only inequalities between values are considered and equal values are neglected. In practical, small random perturbations are added to break equalities. For some signals with quite a lot of equal values, e.g. heart rate variability (HRV) sequence, it is inaccurate to characterize the complexity of this signal by means of PE. Bian and his colleague [37] mapped all the equal values in an epoch into the same symbol as the equal value with the smallest index in this epoch. Results on HRV indicate that this scheme performs better for distinguishing HRV signals.

In this paper, we will introduce a new approach to measure the complexity of time series, termed increment entropy (IncrEn). We focus on the increments of signal as the increments of signal indicate characteristics of dynamic changes hidden in the signal. The basic idea is that each increment is mapped into a word with length of 2 letters, with a sign and its size coded in this word. Then we calculate the symbolic entropy of these words [38-42]. This scheme entails IncrEn an ability to detect either energetic change or structural changes. Moreover, it is of conceptual simplicity. We finally provide demonstrations of its various applications, and the tests on synthetic and real-world data indicates its effectiveness.

The remainder of this paper is organized as follows. Section 2 addresses the notion of increment entropy; Section 3 shows its performance on synthetic trajectories and comparison with PE and AnEn (or SampEn); Section 4 shows its performance on real-world data, such as seizure detection. Section 5 includes the discussion and future works. Finally, we conclude our paper in section 6.

## 1 Increment Entropy

Consider a finite time series $\{x(i)\} | 1 \leq i \leq N$, and construct an increment time series $\{v(i)\}$ from $\{x(i)\}$, where $v(i) = x(i+1) - x(i)$, $1 \leq i \leq N-1$, $N$ is the length of time series. Let m be the length to be compared, we can construct $N-m$ vectors of m dimension from the increment time series, $V(i) = [v(i), v(i+1), \cdots, v(i+m-1)]$. Then we map each element in the constructed vector into a word with a length of 2, where the sign of increment (positive, negative, level) is denoted as a letter, and the size of increment compared with other increments in the same vector is denoted as a letter. As a result, each vector is mapped into a word $\psi = U_{k=1}^{m} s_k q_k$, where $s_k = \text{sgn}(v(k))$, $q_k$ is a quantified value relative to other increments in the same vector. Let $Q(\psi)$ denote total amount of any unique pattern that occurs in the time series, thus its relative frequency is

$$p(\psi) = \frac{Q(\psi)}{N - m + 1} \quad (1)$$

The increment entropy of order $m \geq 2$ is defined as

$$H(m) = -\sum_{i=1}^{(2R+1)^m} p(\psi_i) \log p(\psi_i) \quad (2)$$

where $R$ is the quantifying resolution.

As usual, log is with base 2, thus $H$ is given in bit. In general, $H(m)$ is normalized by $m-1$. Obviously, $H(m)$ is bounded in $[0, (2K+1)^m]$. For example, for a series with ten values $(3, 3, 2, -8, -5, 4, 20, 10, 11, 8)$, we can generate an increment sequence

$(0,1,-10,3,9,16,-10,1,-3)$ from the given series by a simple formula $x_i - x_{i-1}$, $i \leq 2$. We take $m = 2$ and $R = 4$, then we get eight words, each of which is composed up of $s_1 q_1 s_2 q_2$ as follows.

Table I. Symbolization process

| Vectors | $s_1q_1s_2q_2$ |
|---|---|
| 3, 3, 2 | 0, 0,-1,4 |
| 3, 2,-8 | -1, 0,-1, 4 |
| 2,-8,-5 | -1, 4, 1, 1 |
| -8,-5, 4 | 1, 2, 1, 4 |
| -5, 4,20 | 1, 2, 1, 4 |
| 4,20,10 | 1, 4,-1, 3 |
| 20,10,11 | -1, 4, 1, 0 |
| 10,11, 8 | 1, 2,-1, 4 |

In this illustration, the size of increment is quantified in terms of the standard deviation of increment epoch and a coefficient. The word 1,2,1,4 has a replicated copy and its corresponding frequency is $2/8$. The frequency of the remaining six unique words is $1/8$. As a result,

$$H(2) = -\frac{1}{8}\log(\frac{1}{8}) \times 6 - \frac{2}{8}\log(\frac{2}{8}) \approx 1.9062 \quad (3)$$

$$h(2) = \frac{H(2)}{2-1} \approx 1.9062 \quad (4)$$

## 2 Simulation and results

We conducted a variety of simulations both on chaotic signals and on regular signals to examine IncrEn. We compared IncrEn with PE and SampEn on these synthetic data in various aspects, such as reliability for detecting abrupt change, data length effect, and stability.

### 2.1 Results from chaotic time series

First, we demonstrate that IncrEn can effectively and accurately indicate the complexity of dynamical system using the well-known logistic map $x_n = r \times x_{n-1} \times (1 - x_{n-1})$ already shown in previous studies [25,33]. We generated 20,000 values of the logistic map with $3.5 \leq r \leq 4$ and initial $x_0 = 0.303$. The bifurcation diagram of the logistic map is shown in Fig.1a. We can clearly see from Fig.1b-d that IncrEn is capable of distinguishing the chaotic degree, similar to the positive Lyapunov exponent. This is completely consistent with the results in previous studies [25, 33]. We can obviously see from Fig.1b-d that the functions $h_4$, $h_6$ and $h_8$ are very similar. This justifies that IncrEn of lower order is as IncrEn of higher order when appling to real-world data [25].

We further examine the effects of Gaussian observational noise on IncrEn. Comparing Fig.1d with Fig.1c, it is obvious that IncrEn with Gaussian noise is entirely identical to that without Gaussian noise. As shown in Fig.1d, IncrEn, $h_8$, only grows up slightly, compared to that without noise, even in low-period window. This is very distinct from PE as PE is relatively larger in the low-period window, e.g. period-4 and period-3 windows (see ref [25] Fig.2e).

We calculate the mean and variance of the IncrEn of data with various length through taking 1000 time series produced by the logistic map with $r = 4$ to address the effect of data length N. In Fig. 3a $h_m$ increases at the beginning, reaches the highest at $m = 4$, then decreases. The variance of $h_m$ is rather small as shown in Fig.3b. In general, a smaller data size compared to $(2R+1)^m$, will cause a bias in IncrEn. In practice, we recommend the order $m = 2,3,4,5$ and the quantifying resolution . $R < 4$.

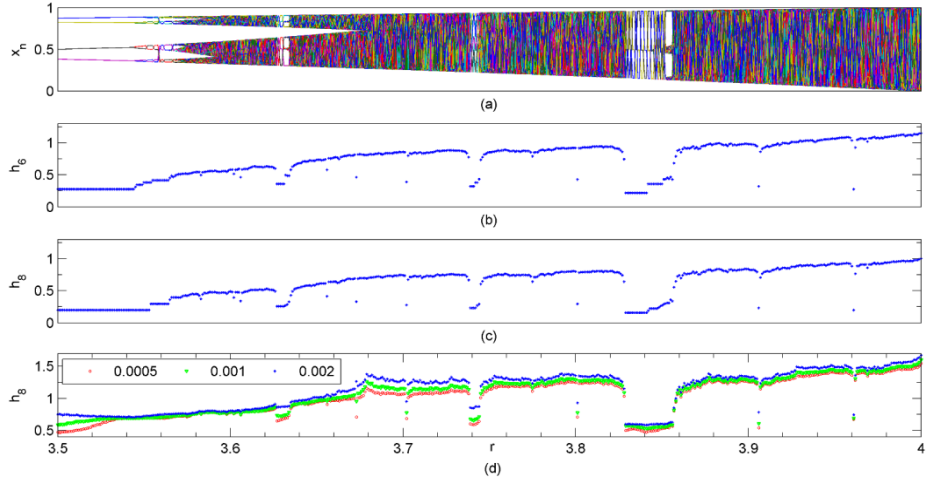

Figure 1. Logistic equation for varying control parameter r and corresponding IncrEn with varying scale. a) Bifurcation diagram, b) Increment Entropy, $h_6$, c) $h_8$, d) $h_8$ with Gaussian observational noise at standard deviations.

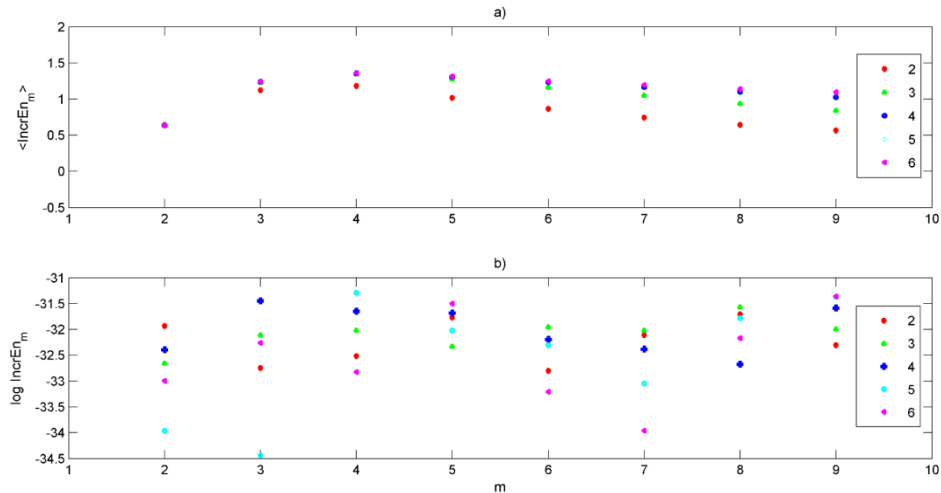

Figure 2. Order m choice and data length effect. a) Mean $\langle h_m \rangle$ of logistic map (r = 4) with $10^k$ data points (k = 2, 3, 4, 5, 6). Horizontal: embeding dimension m, b) Corresponding standard deviation $\sigma$ of $h_m$.

## 2.2 Comparison with PE and SampeEn

The notion of increment entropy is motivated by the concept of permutation [25] and approximate entropy [16], thus we examined the relationship between IncrEn and PE and SampEn to illustrate its features and applications. The number of possible formed patterns $(2R+1)^m$ produced in IncrEn is roughly more than that produced in PE, $m!$, for smaller order m (it is also dependent on quantifying resolution too). Thus IncrEn is much larger than PE for random or chaotic signals, while it is a little lower than PE for regular signals because less patterns are generated from the computation based on increments of time series. For regular signals, the distances between vectors are commonly smaller than the specified threshold, which leads to a higher conditional probability. As a result, SampEn (or ApEn) is lower than IncrEn for regular time series.

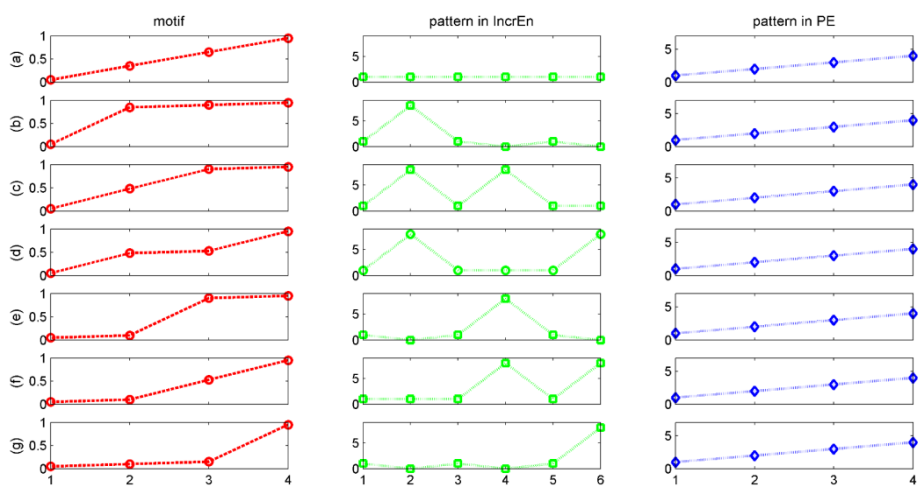

Figure 3. Symbolization of analogous pattern in IncrEn and PE.

### 2.2.1 Distinguishing analogous patterns in IncrEn and PE

The left panel of Fig.4 shows seven patterns that could be attributed to an identical motif. These seven vectors are in the same order, thus all of them are transformed into the same word in PE. In contrast, in IncrEn, they are mapped into seven different words because of the distinct increment pattern in each vector, although they have an identical ordinal pattern. This endows IncrEn a greater sensitivity to changes hidden in time series. However, some of them could be classified in the same category in terms of the distance between each vector. For example, it is clear that the vector presented in Fig.3b is symmetrical to that in Fig.3g with respect to the vector in Fig.3a, thus obviously their distance to the vector in Fig.3a is equal. For appropriate threshold $r$ the three vectors are deemed identically in SampEn and ApEn even though they have very distinct appearances.

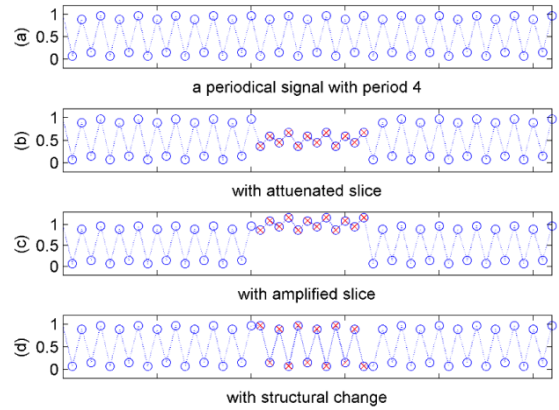

Figure 4. Detection of Energy and Pattern Change. (a) Regular time series consists of 300 identical slices [0.07, 0.89, 0.15, 0.97]. (b) Time series (a) interspersed with 3 slices [0.37, 0.59, 0.45, 0.67]. (c)mTime series (a) interspersed with 3 slices [0.37, 0.59, 0.45, 0.67] adding 0.5 offset. (d) Time series (a) interspaced with 3 slices [0.97, 0.15, 0.89, 0.07].

2.2.2 Detecting energetic change and structural change

The synthetic data with impulsive spikes are usually used to illustrate the failure of PE for detecting abrupt change in amplitude in previous studies (see [33] Fig.1, [36] Fig.2a). Either spike or burst in signal is one kind of energetic change in signal. In this paper, we demonstrated the performance of increment entropy on an entire regular time series and regular data interspersed either with a few amplitude amplified epochs (AMP.) or with a few amplitude attenuation epochs (ATT.), as shown in Fig.4a-c. To demonstrate the performance of IncrEn for detecting structural change, we synthesized a regular signal interspersed with a few reverse ordinal epochs (Struct.) too (Fig.4d).

Table 2. Entropy of regular signals with and without distorted slices.

| Entropy | | Regular | ATT. | AMP. | Struct. |
|---|---|---|---|---|---|
| IncrEn | m=2 | 0.6931 | 0.7328 | 0.7345 | 0.7345 |
|  | m=3 | 0.3466 | 0.3694 | 0.3698 | 0.3698 |
|  | m=4 | 0.2310 | 0.2477 | 0.2479 | 0.2479 |
| PE | m=2 | 0.6930 | 0.6930 | 0.6930 | 0.6930 |
|  | m=3 | 0.6932 | 0.6932 | 0.6932 | 0.6932 |
|  | m=4 | 0.4622 | 0.4661 | 0.4650 | 0.4650 |
| SamEn | | 0.0000 | 0.0000 | 0.0000 | 0.0008 |

We calculated IncrEn, PE, and SmapEn of each signal presented in Fig: 4 and the results in Table 2. As mentioned in the previous section, overall, IncrEn is smaller than PE in various order $m = 2,3,4$. It is clear that IncrEn is able to discriminate all three signals with distorted epochs from regular one in all three orders ($m = 2,3,4$). In contrast, PE can not distinguish distorted signals from regular ones because PE of regular signal is equal to those with distorted epochs in lower order, $m = 2,3$, except for a small difference in order 4. For all four signals, sample entropy is zero or approximate to zero, which suggests that sample entropy fails to detect these changes, energetic or structural. Compared to PE and SampEn, IncrEn is sensitive to a fraction of subtle changes although the differences between distorted and regular conditions are smaller. The case for higher order m = 5 is similar to that for order $m = 4$, for IncrEn and for PE. This suggests that multi-scale entropy can reveal different aspects of data relative to single-scale entropy. However, we have to point out that the results may be biased for data with short length when using multi-scale entropy.

2.2.3 Stability of IncrEn, PE and SampEn

Figure.5

As presented in the previous section, the variability of IncrEn is rather small, which is comparable to the performance of PE [25]. We further computed the variability of IncrEn, PE, and SampEn on a random time series, a Gaussian noise. As shown in Fig.5, IncrEn is universally much higher than PE for random time series. The track of IncrEn is approximately of the same plain as that of PE (variance $< 10^{-3}$ and $10^{-6}$ for data length W = 500 and 1000). In contrast, the track of SampEn is full of fluctuations (variance = 0.0337 and 0.0092, for W = 500 and 1000) even though the data length is much longer than 100, which is required for the calculation of SampEn [8]. This indicates that IncrEn performs as well as PE, and both of them are much better than SampEn in terms of their variability.

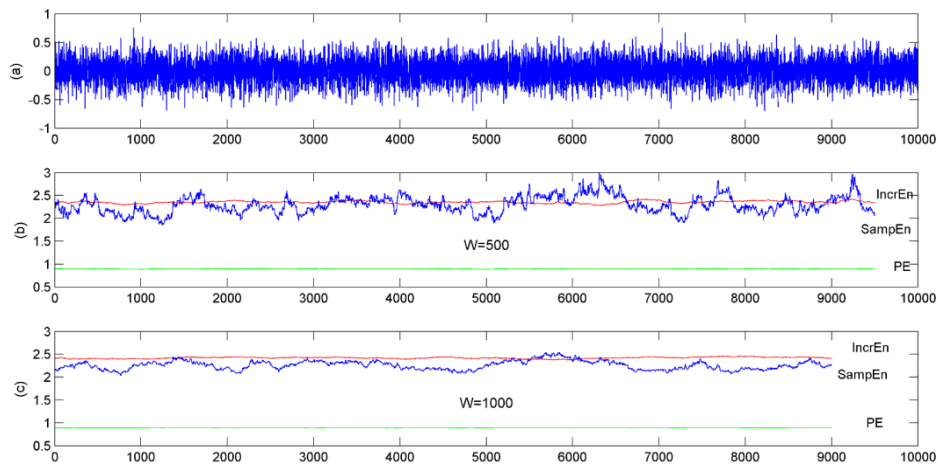

Figure 5. Time-dependent IncrEn, PE and SampEn of random noise (Data length = 20,000, Window size = 200).

## 2.3 Application to seizure detection from EEG signals

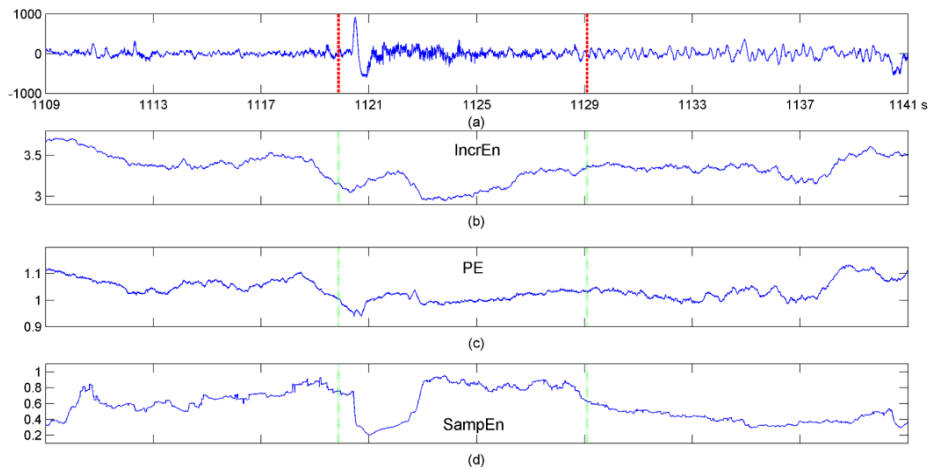

Figure 6. Detecting seizure of a short time from epileptic EEG signals. (a) A part of epileptic EEG of a female patient aged 7 years. It contains a seizure of and is recorded from Fz-Cz electrode with a sample rate of 256 Hz. (b) IncrEn. (c) PE. (d) SampEn. All of them are computed using a sliding window of 500 samples with 499 samples overlap.

Entropy has been an effective approach to be used to detect seizure from EEG signals [27, 33, 36]. We calculate IncrEn of epileptic EEGs that contain a seizure and compare it with PE and SampEn. The data set is continuous scalp EEG sampled at 256 Hz, and data sets are available from the CHB-MIT database [43, 44]. The earliest EEG changes associated with each seizure are indicated by a clinical expert [43, 44]. We choose two different data sets to evaluate IncrEn for detecting dynamical changes in real time series. One contains a long seizure of 120 seconds and the other contains a short seizure of 10 seconds. The results for detecting a short seizure are presented in Fig.6. A seizure in EEG signal begins at 1120s and ends at 1129s (see Fig.6a). It is clear that IncrEn drops when the seizure occurs and remains yet much lower during seizure than before and after seizure (see Fig.6b). IncrEn reaches the minimum at around 1123s. This is consistent with previous study that Entropy descends during seizure stage and increases when seizure ends [33, 45]. In contrast, PE becomes almost level for a long time after 1123s, irrespective of seizure (see Fig.6c). SampEn falls sharply at initial phase of seizure, and rises at around 1123s. In comparison to PE and SampEn, the decrease of IncrEn overall captures the whole process of seizure, rather than only the initial part of seizure, and moreover, the drop corresponding to seizure is more evident.

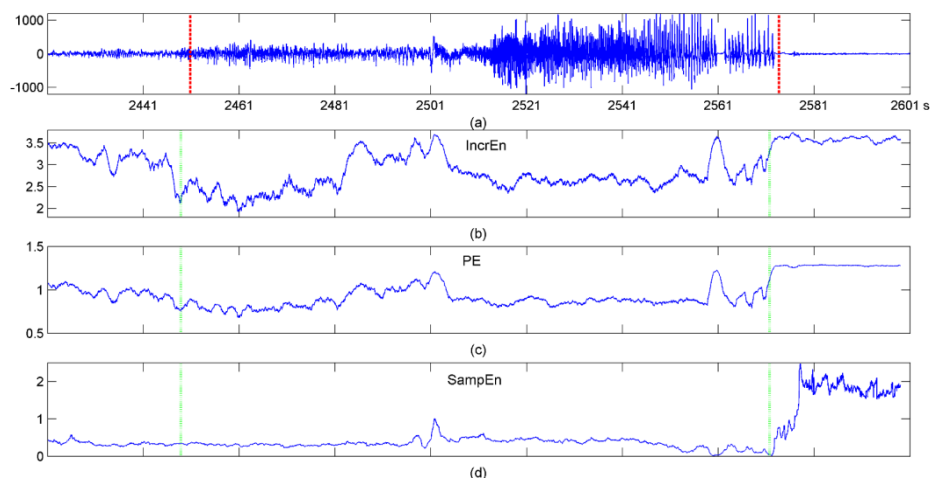

Figure 7. Detecting seizure of a long time from epileptic EEG signals. (a) A part of epileptic EEG of a male patient aged 16 years. It contains a seizure of and is recorded from Fz-Cz electrode with a sample rate of 256 Hz. (b) IncrEn. (c) PE. (d) SampEn. All of them are computed using a sliding window of 500 samples with 499 samples overlap.

Figure.7

We calculate IncrEn of a long seizure as well. Unlike the above short seizure, there are a few transient pauses during a long seizure as shown in Fig.7a. Similar to the case of a short seizure, IncrEn goes downwards obviously at the initial phase of seizure and goes upwards at the end of seizure. Further, it is evident from Fig.7b that IncrEn is sensitive to these transient pauses within seizure, which means IncrEn can accurately track the process of seizure. So is PE, except that the drop at the initial phase is much smaller compared with IncrEn. Although it is far lower during seizure than after seizure, SmapEn fails to indicate these transient pauses within seizure. Together with results of a short seizure, IncrEn can not only accurately detect seizure in EEG signals, but also sensitively capture various changes within seizure.

## 3 Discussion

To date, there is not a universal definition of complexity of time series. Most of measures of complexity monotonically increase with randomness, giving a lower value for regular sequences but the maximum for random sequences. Other definitions assign a lower value for both completely regular and pure random sequences, but give the maximum value for the intermediate stage [18, 46-48]. However, from the point of view of predictability, neither chaotic sequence nor random sequence is predictable. Both of them are highly complex [13]. Our approach is an intuitive notion of complexity in that IncrEn grows up with increasing disorder, maximizing at random stage but minimizing at regular stage. On this respect, it is the same as PE and SampEn. IncrEn is lower for a regular sequence, higher for a random one, and intermediate for a chaotic one, thus is a measure of the degree of sequence predictability.

Our simulation has demonstrated that IncrEn is more sensitive not only to the subtle alternation of amplitude with identical structure, but to tiny modification of structure or pattern, as shown in Fig. 2 and Fig. 3. This allows for accurately capturing the subtle changes hidden in sequences, irrespective of whether energetic or structural. As a result, IncrEn can be used to extract various features hidden in time series and to detect dynamical change in time series relating to internal or external factors.

In recent years, entropy has been an effective method to characterize and to distinguish the physiological signals across diverse physiological and pathological conditions, such as HRV abnormality [20, 21], neural disorder [22, 49, 50], gaits analysis [23], etc. As shown in Fig.6 and Fig.7, IncrEn exhibits better performance on seizure detection from real epileptic EEG signals than PE and SmapEn. A prominent merit of IncrEn lies in that the computation of IncrEn does not make any assumption on data, which makes it applicable to diverse signals. In particular, it is an appropriate method to examine the characters of those signals consisting of most equal values, e.g. HRV, equal values in signals can be coded as equivalently as those inequalities, unlike PE that does not consider the equal values [25, 37].

IncrEn exhibits robust to observational noise. As shown in Fig. 1e, the noise do not disturb the appearance of IncrEn, even in lower-period window where PE is higher and flips the case without noise (see Fig.2e in ref [25]). The noise only causes a slight increase in IncrEn. Overall IncrEn with noise is consistent with that without noise (comparing Fig.1d and Fig.1e). It suggests that IncrEn is suitable to analyzing noisy data.

The data length has a great impact on entropy computation [8, 16, 24, 25, 51]. Even though the data length is 100, the variance of IncrEn across different data length is small as shown in Fig.2. We calculated the IncrEn, PE and SmapEn for a Gaussian noise sequence for the sake of comparison. Both IncrEn and PE have quite a lower variance. In contrast, variance of SampEn is much higher as shown in Fig. 5, although SampEn is largely independent of record length [8, 24]. This indicates that IncrEn is powerful tool for analyzing very short signals.

## 4 Conclusion

In this paper we present a novel measure to quantify the complexity of time series, as well as to characterize and distinguish the time series across regular, chaotic and random states. Results of a series of simulations and real data have illustrated the powerful function and the potential applications of IncrEn to reveal the characteristics of a variety of signals and to discriminate sequences across various conditions. Our approach has a wider dynamical range compared to the established PE and is more reliable than sampEn or ApEn. It is a promising method to characterize time series of diverse fields.